\documentclass[showpacs,preprintnumbers,amsmath,amssymb]{revtex4}

\usepackage{graphicx}
\usepackage{bm}

\begin{document}

\title{Universality class of S=$\frac{1}{2}$ quantum spin ladder system
with the four spin exchange}
\author{Keigo Hijii and  Kiyohide Nomura}
 \affiliation{\it Department of Physics,Kyushu University, Fukuoka 812-8581, Japan}
\date{\today}

\begin{abstract}

We study s=$\frac{1}{2}$ Heisenberg spin ladder with the four spin exchange.
Combining numerical results with the conformal field theory(CFT)
\cite{gins}, 
we find a phase transition with central charge c=$\frac{3}{2}$.
Since this system has an SU(2) symmetry, 
we can conclude that this critical theory is described
by k=2 SU(2) Wess-Zumino-Witten model with Z$_2$ symmetry breaking.

\end{abstract}


\maketitle

\section{INTRODUCTION}

 Quantum spin ladder systems \cite{sci} have been studied in relation
with high-T$_c$ superconductivity and 
Haldane's conjecture \cite{haldane}.
These studies have been done mainly with 2-body interaction terms.
Recently from experiments on dispersion curve \cite{exp1}, \cite{exp2},
it is suggested that 4-spin exchange interactions play an important role.

In general, 
spin interactions originate 
from electron exchange interactions.
It is well known that 
Heisenberg model is derived from
the second order  
perturbation expansion 
in strong coupling limit of Hubbard model.
Similarly higher order terms 
give many body spin interactions.
For example,
fourth order terms give the four body spin interactions, 
e.g. 
$\left( \vec{S}_i \cdot \vec{S}_j \right) \left( \vec{S}_k \cdot \vec{S}_l \right)$.
In fact,  in the one dimensional (1D) two leg ladder,
many body spin interactions appear from electron exchanges
\cite{Takahashi},\cite{Takahashi2}. 
Note that 1D nearest neighbor Hubbard model does not
give many body interactions,
because the cyclic path is impossible on this lattice.
Therefore, it is important to understand 
the spin ladder systems
including 4-spin exchange term.

Apart from the spin ladder systems,
many-body interactions appear in 2D, 3D systems.
It is experimentally known that
many-body interaction can not be neglected
especially in $^3$He on graphite \cite{3he}, 
and Wigner crystal \cite{wigner}. 
As another example, 
in high T$_c$ superconductor described by 2D Hubbard model,
many-body spin interactions appear and 
they may be important physically \cite{ueda}.

This paper is organized as follows.
In the next section, we shortly review for $s=1/2$ two-leg
spin ladder systems.
In Sec. 3, we discuss the phase transition and the universality class
from numerical results and the conformal field theory.
In Sec. 4, we summarize results.
In appendix, we briefly explain theories used in this study, that is,
$s=1$ solvable models, WZW model, logarithmic corrections.

\section{OVERVIEW OF SPIN LADDER SYSTEMS} 
\begin{figure}
 \includegraphics[width=10cm]{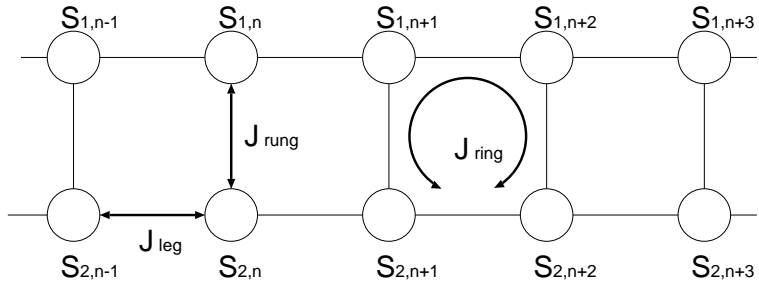}
 \caption{Two-leg spin ladder of eq. (1) 
 }
\label{fig:jring-fig}
\end{figure}

We treat the following Hamiltonian
\begin{equation}
  H = J_{leg} \sum_{i=1}^{L} \sum_{\alpha=1,2} \vec{S}_{\alpha,i} \cdot 
      \vec{S}_{\alpha,i+1} +
     J_{rung} \sum_{i=1}^{L} \vec{S}_{1,i} \cdot \vec{S}_{2,i} 
      + J_{ring} \sum_{i=1}^{L} \left( P_{i,i+1} + P^{-1}_{i,i+1} \right),
\end{equation}  
where $J_{leg}$ is the intrachain coupling,
$J_{rung}$ is the interchain coupling, 
and $P_{i,j,k,l}$ is the four spin exchange term
which is written with spin operators
\begin{eqnarray}
P(i,i+1)+P^{-1}(i,i+1)  = & & \frac{1}{16} +
 \vec{S}_{1,i}{\cdot}\vec{S}_{1,i+1}+\vec{S}_{2,i}{\cdot}\vec{S}_{2,i+1}+\vec{S}_{1,i}{\cdot}\vec{S}_{2,i+1} \nonumber \\
& & +\vec{S}_{2,i}{\cdot}\vec{S}_{1,i+1}+\vec{S}_{1,i}{\cdot}\vec{S}_{2,i}+\vec{S}_{1,i+1}{\cdot}\vec{S}_{2,i+1} \nonumber \\
& & +4{\{}(\vec{S}_{1,i}{\cdot}\vec{S}_{2,i})(\vec{S}_{1,i+1}{\cdot}\vec{S}_{2.i+1}) \nonumber \\
& & +(\vec{S}_{1,i}{\cdot}\vec{S}_{1,i+1})(\vec{S}_{2,i}{\cdot}\vec{S}_{2,i+1}) \nonumber \\
& & -(\vec{S}_{1,i}{\cdot}\vec{S}_{2,i+1})(\vec{S}_{2,i}{\cdot}\vec{S}_{1,i+1}){\}},
\end{eqnarray}
and indexes $i$ are for longitudinal direction, 
$\alpha$=1,2 are indexes of spin chains,
(see Fig.\ref{fig:jring-fig}).
In this section, 
we shortly review the model and related studies.

Quantum spin ladder systems have been well studied.
Recently several quantum spin ladder materials 
are found out experimentally.
and they attract much attention \cite{sci},\cite{da}. 
Theoretically, 
they have been studied with bosonization method and 
numerical calculations.
For only 2-body interaction, there are many studies.
See for example, 
\cite{sci},\cite{snt},\cite{ks},\cite{kss},\cite{da},
and references therein.

When there is the four spin exchange, 
it is expected that strange phase transitions
may appear.
In this region, relatively there are few studies.
Brehmer {\it et al}. numerically calculated the dispersion curve 
in the s=$\frac{1}{2}$ spin ladder 
with the four spin exchange \cite{bmmnu}.
They discussed it from the view point of the symmetry breaking.
Sakai {\it et al.} \cite{sh} studied with numerical calculation for  
the system with the magnetization plateau.
In that paper, they found out that this system is 
Tomonaga-Luttinger (TL) liquid with central charge $c=1$, 
when magnetization m=$\frac{1}{2}$,
and that BKT transition occurs between TL phase and 
plateau phase region. 
Honda {\it et al.} \cite{hh} investigated the energy gap and 
the spin-spin correlation function 
with the density matrix renormalization group method.
Recently experimentists found the real material with 4-spin 
cyclic exchange,
i.e., La$_6$Ca$_8$Cu$_{24}$O$_{41}$.
In that experiment, the dispersion relation was observed by
neutron scattering 
,which suggests four-body terms \cite{exp1},\cite{exp2}.

When J$_{rung}<0$,
s=$\frac{1}{2}$ quantum spin two-leg ladder systems  
can be mapped to s=1 systems.
Thus they can be considered as a generalization of Haldane's conjecture 
\cite{haldane} for ladder systems.
Haldane phase is characterized by the string order parameter,
which is related to a hidden $Z_2 \times Z_2$ symmetry breaking.
In s=$\frac{1}{2}$ ladder systems with four spin exchanges, 
the string order parameter was discussed by Fath {\it et al} \cite{fls}.
In that paper, there are two type gapped phases.
One is a rung singlet type, another is the AKLT type \cite{aklt}.
Using density matrix renormalization group, Legeza {\it et al}. 
\cite{lfs} pictured phase diagram in these systems.
But they have not decided phase boundary properly.
Nersesyan {\it et al}. \cite{nt} referred to the relation between 
s=$\frac{1}{2}$ ladder system with four spin exchange 
and s=1 quantum spin chain with bilinear-biquadratic term
(see Appendix A). 
In that paper, they examined behaviors of a correlation function of  
relative staggered magnetization and 
a correlation function of relative dimerization field,
and suggested the phase transition with 
central charge $c=\frac{3}{2}$.

When $J_{rung}<0$, we can relate 
s=$\frac{1}{2}$ quantum spin ladder system with four spin exchange
to s=1 quantum spin chain with bilinear-biquadratic term.
When $J_{rung}>0$, is it obvious ?
The discussion of Legeza {\it et al}. \cite{lfs} is different from that of
 Nersesyan {\it et al}. \cite{nt}.
Legeza {\it et al}. \cite{lfs} wrote that this phase transition is between 
a nondegenerate singlet state and a fourfold-degenerate ground state.
The fourfold-degenerate ground state realizes 
because the singlet-triplet
gap disappear. 
Nersesyan {\it et al}. \cite{nt} wrote that,
there is the region where 
the triplet Majorana excitation is lower than the singlet one,
in which it is enough to consider only the triplet Majorana excitation,
thus the phase transition belongs to the second order type 
with $c=\frac{3}{2}$.
Although both of them predicted Takhtajan-Babujian type transition
point ($c=3/2$) in the $s=1/2$ spin ladder (with four-body terms),
there is a difference on other critical phenomena between them.
Nersesyan {\it et al}. \cite{nt} referred to $c=1/2$
(the singlet Majorana field becomes massless  in another region),
Legeza {\it et al}. \cite{lfs} referred to $c=2$ 
Uimin-Lai-Sutherland model \cite{uls}.
There remains controversial in this region.

\section{RESULTS AND DISCUSSION}

We study $J_{leg}$=$J_{rung}$=1 case with varying $J_{ring}$
coupling in the system (1) with no magnetizations,
combining numerical calculation and conformal field theory(CFT).

\subsection{numerical results and universality discussion}

\begin{figure}
 \includegraphics[width=10cm]{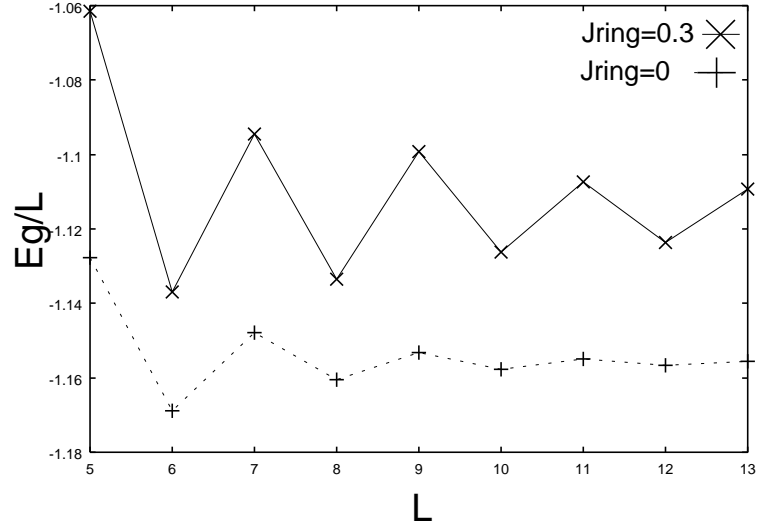}
 \caption{
 Size dependence of the ground state energy 
at $J_{ring}=0,0.3$. 
 }
 \label{fig:siz-fig}
\end{figure}

\begin{figure}
 \includegraphics[width=10cm]{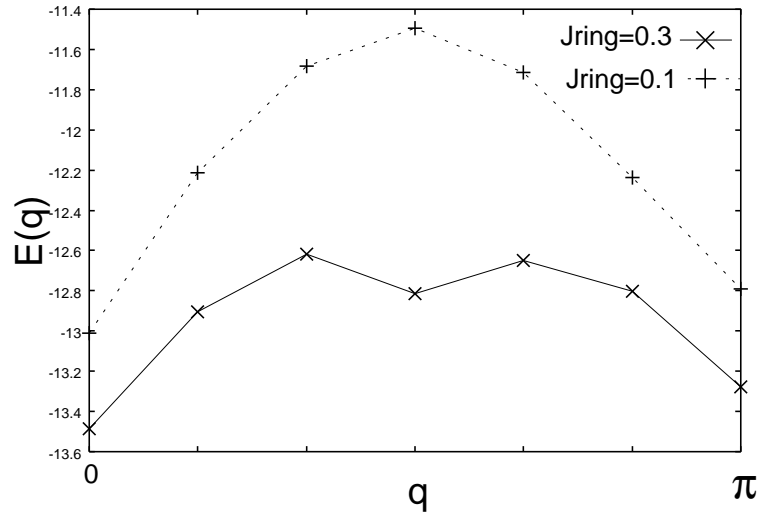}
 \caption{
 Dispersion curve at J$_{ring}$=0.1 and J$_{ring}$=0.3for the system size L=12 (N=24).
 }
 \label{fig:disp24-fig}
\end{figure}

At first, we study the size dependence of the ground state energy,
in order to see the type of periodicity ($Z_2$ or $Z_3$)
under periodic boundary condition (PBC). 
In Fig.\ref{fig:siz-fig},
we can see that the ground state energy per site $E_g (L) /L$ 
oscillates according to $L$ even or odd (even cases are lower).
This suggests that translational invariance by one site
may be spontaneously broken.
For another evidence of $Z_2$ symmetry breaking,
we study the dispersion curve,
which shows that the lowest excitation has soft modes at $q= \pi$
(see Fig.\ref{fig:disp24-fig}).
Thus we will study the system size L=N/2=6,8,10,12.

In order to discuss what type of phase transitions may occur,
we use the effective central charge $c$,  
obtained from the finite size scaling of the ground state energy.
From the conformal field theory at the critical point,
the finite size scaling of ground state energy under 
the periodic boundary condition (PBC) is
the following \cite{bcn},\cite{a},
\begin{equation}
 E_g(L) = \epsilon  L - \frac{ \pi  v  c}{6  L},
\end{equation} 
where $ E_g(L)$ is the ground state energy for size $L$,
$\epsilon$ is the energy per site in the infinite limit,
$v$ is the spin wave velocity.

In addition to the ground state energy,
we should obtain $v$ numerically to decide $c$ in equation (3).
The spin wave velocity $v$ is obtained from
\begin{equation}
 v(L) = \frac{L}{2 \pi} \left[ E \left(q=\frac{2 \pi}{L} \right) - E(q=0
) \right],
\end{equation} 
where $q$ is a wave number.
Then we extrapotale $v(L)$ as

\begin{equation}
 v \left( L \right) = v_{} + a \frac{1}{L^{2}}
 + b \left( \frac{1}{L^{2}} \right)^{2} + \mbox{higher order}.
\end{equation}

\begin{figure}
 \includegraphics[width=10cm]{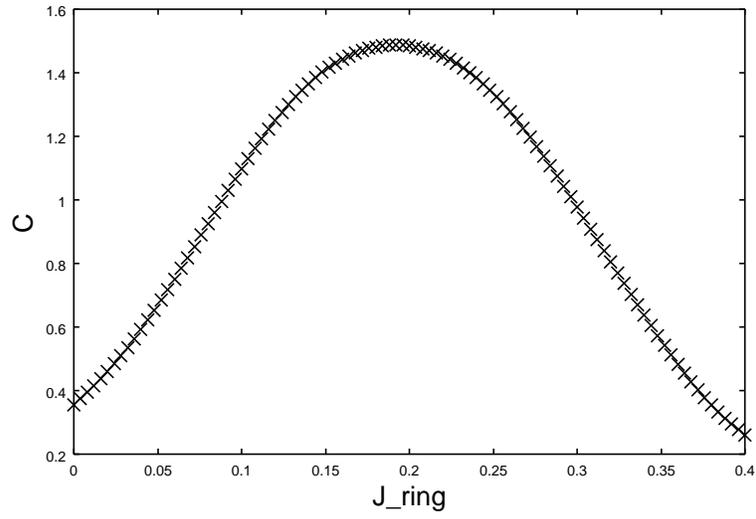}
 \caption{
 Effective central charge
 }
 \label{fig:cc-fig}
\end{figure}

In Fig.\ref{fig:cc-fig}, we show the obtained effective central charge.
The effective central charge shows a maximum value 
$c$=1.49 at $J_{ring}$=0.192.
According to Zamolodchikov's c-theorem \cite{kz}, 
the effective central charge shows maximum at the infrared fixed point.
Thus we can conclude that there is a phase transition 
with the central charge
$c=\frac{3}{2}$.

Only from the central charge, there remain several possibilities 
for universality class.
But considering the symmetry of the system,
we can discuss in detail. 
Combining the overall SU(2) symmetry of Hamiltonian,
we can relate central charge $c=\frac{3}{2}$ in conformal field theory 
to the topological coupling constant 
(Kac-Moody central charge) $k$ in SU(2) Wess-Zumino-Witten model.
The relation between $k$ and $c$ is given by
\begin{equation}
 c = \frac{3k}{k+2}.
\end{equation}
Thus the critical theory c=$\frac{3}{2}$ with SU(2) symmetry 
is described 
as the $k=2$ SU(2) Wess-Zumino-Witten model.
This universality class is characterized with scaling dimensions
$x=\frac{3}{8}$ (parity odd,$q=\pi$), $x=1$ (parity even,$q=0$) for
the primary field (in Kac-Moody algebra), 
$x=2$(parity even,marginal,$q=0$) for the (Kac-Moody) descendant field
(see appendix B).
They mean that a second order transition occurs
 with $Z_2$ symmetry breaking.
And the mass (energy) gap is generated near the critical point
as $\Delta E \propto | J_{ring}-J_{ring}^c|$ from $x=1$ term.
Here the mass gap $\Delta E$ is related with the correlation
length $\xi$ as $\Delta E \propto \xi ^{-1}$.

Next we study lowest excitations with wave number $q=\pi$,
which consist of $s=0$ (singlet) and $s=1$ (triplet).
We can expect that these excitation are related with the scaling
 dimension ($x=\frac{3}{8}$,$q=\pi$) in the $k=2$ SU(2) WZW model.
In general, from CFT, scaling dimension can be calculated from
the excitation energy for finite size under PBC as \cite{c}

\begin{equation}
 \Delta E_{i} = E_i - E_0 = \frac{2 \pi v}{L} x_i.
\end{equation}

\begin{figure}
 \includegraphics[width=10cm]{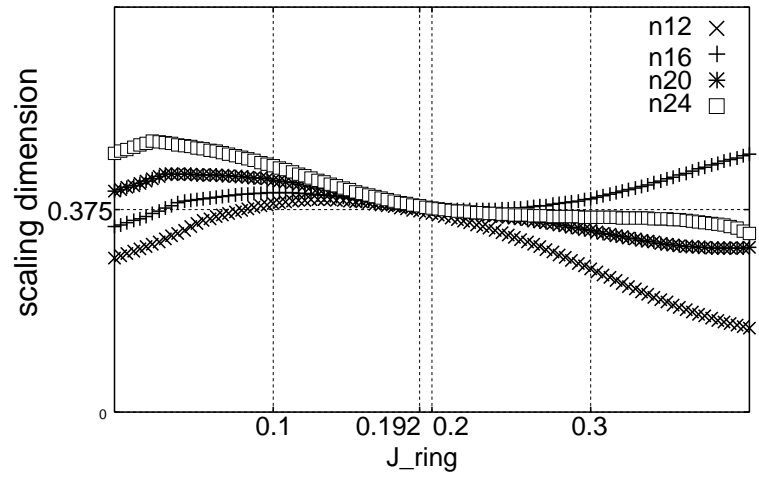}
 \caption{
 Scaling dimension $x$ at $q=\pi$.
 }
 \label{fig:sdim-fig}
\end{figure}

Unfortunately, in the $k=2$ SU(2) WZW model, there are marginal 
corrections, 
1/$\ln L$, which are difficult to treat.
Thus we should remove them with the method in appendix C. 
After removing logarithmic corrections,
scaling dimension is almost independent on the system size 
at critical point $J_{ring}=0.192$ (See Fig.\ref{fig:sdim-fig}).
Therefore, we can conclude that this universality class belongs to
the $k=2$ SU(2) WZW type.

Finally, we comment on the effect of the marginal operator 
to the ground state energy and the velocity.
For ground state energy, it is proved that,
\begin{equation}
E_g \left( L \right) = \epsilon L 
  - \frac{\pi v}{6 L} 
    \left( c + \frac{b}{\left( \ln L\right)^3} \right)
\end{equation}
where $o\left( \frac{1}{ \ln L} \right)$, 
$o\left( \frac{1}{\left( \ln L\right)^2} \right)$ terms 
do not exist \cite{c2}.
We have neglected the $o\left( \frac{1}{\left( \ln L \right)^3} \right)$ 
contribution
(which is small enough) to obtain the effective central charge.
For the velocity, it is proved that there is not 
$o\left( 1 / \ln L \right)$ term in current-current correlation
\cite{tnhs}.
Therefore, we have neglected 
logarithmic correction in equation (5).

\subsection{symmetry breaking and boundary condition}

After studying the universality class,
we proceed to discuss the symmetry breaking and the order parameter.
Although finite size effect prevents us from drawing a definite 
conclusion, we can consider several clues relating to the 
ordered state.

\begin{figure}
 \includegraphics[width=10cm]{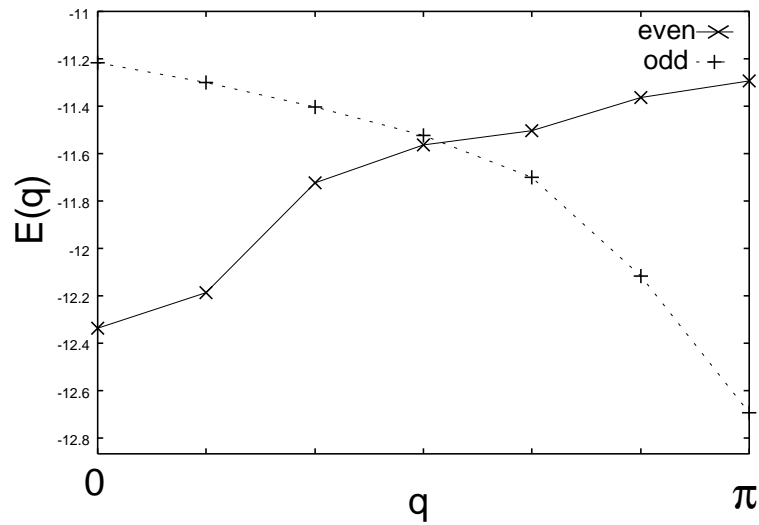}
 \caption{
 L=12 (N=24) dispersion curve with rung parity
 at $J_{ring}=0.15$. 
 }
 \label{fig:24-fig}
\end{figure}

At first, for $J_{ring} < J_{ring}^c$, the triplet excitation
is lower than the singlet excitation at $q=\pi$,
whereas for $J_{ring} > J_{ring}^c$, the singlet excitation 
is inclined to be under the triplet excitation.
But this is not clear in terms of the finite size effect.
Secondary, besides the translational symmetry breaking, 
it is observed a rung-parity symmetry breaking,
reversing each spin chain ($S_{1,i} \leftrightarrow S_{2,i}$),
which
Brehmer {\it et al.} discussed on the dispersion curve.
We confirmed their dispersion curves at $J_{ring}=0.15$
(see Fig.\ref{fig:24-fig}).
In summary, in the $L \rightarrow \infty$ limit under PBC,
it is expected that, for $J_{ring} > J_{ring}^c$,
the ground state ($q=0, S_{z}^{tot}=0, P_{rung}=even$)
will be degenerate with the lowest state
($q=\pi, S_{z}^{tot}=0, P_{rung}=odd$),
while for $J_{ring} < J_{ring}^c$,
there is no symmetry breaking.

Honda {\it et al.} studied the same system 
as ours \cite{hh}.
Using the density matrix renormalization group method,
they calculated the spin gap and spin-spin correlation function.
They claimed that there occurs a phase transition from a massive
phase to a massless phase at $J_{ring} \simeq 0.3$.

Since they studied the open boundary condition,
we think that they saw the edge state.
For the $s=1$ case, in the Haldane phase, 
the edge state can be degenerate
under the open boundary condition
\cite{ke}.
The four lowest energy states are almost degenerate, 
whose energy gap decays $\exp \left( - L / \xi \right)$,
and
there is a Haldane gap between them and the continuum energy spectrum.  
On the finite size effect, the edge state is well defined 
when it is sufficiently localized, that is,
the system size is large enough compared with the correlation length. 
However, since the correlation length becomes very long
near the critical point, so the finite size effect becomes large.
Therefore, considering edge states, 
it is possible to explain 
why Honda {\it et al}. estimated the critical point
$J_{ring}^c$ larger than ours.
Numerically, edge states can be confirmed by investigating 
the ground state energies 
under open boundary condition with higher magnetizations. 
Note that edge states can be detected experimentally.

\section{SUMMARY}

In the present paper, we studied $s=\frac{1}{2}$ quantum spin 
ladder with four spin exchange,
in order to see the phase transition type in this system.

We saw the behavior of the central charge and the scaling dimension,
which have logarithmic corrections.
At first, we calculated the central charge, since its correction
is small.
The central charge has the maximum value $c=1.49$ at $J_{ring}$=0.192.
Removing the logarithmic correction for the scaling dimension,
we obtained $x \approx \frac{3}{8}$ near $J_{ring}$=0.192.
In this region, the scaling dimension do not have a significant 
size dependence.
So we can conclude that 
this phase transition belongs to a $k=2$ SU(2) WZW type.

\section*{ACKONWLEGEMENT}

We thank Prof. Takahashi informing us on the relation between
electron exchange and many body spin interaction.
We acknowledge A. Kitazawa for fruitful discussions 
about spin ladder systems with many body interactions.
We thank to S. Qin, H. Inoue, and S. Hirata in other things.
The numerical calculation in this paper have been done 
based on the package
TITPACK ver 2.0 by H. Nishimori.

\appendix

\section{s=1 quantum spin chain with bilinear-biquadratic term}

We shortly review the following Hamiltonian.
\begin{equation}
   H = \sum_n \left( \mbox{cos} \theta
      \left( \vec{S}_n \cdot \vec{S}_{n+1} \right)
    + \mbox{sin} \theta \left( \vec{S}_n \cdot \vec{S}_{n+1} \right)^2
  \right).
 \end{equation}
This model is well researched.
The phase diagram of this model is known in numerically 
\cite{fs},\cite{xg}.
And this model is exactly solved by Bethe Ansatz,
at the special point $\theta$=$-\frac{\pi}{4}$,$\frac{\pi}{4}$,0.
The critical point $\theta$=$-\frac{\pi}{4}$ 
is Takhtajan-Babujian point \cite{TB} \cite{kn}.
This is the critical point between dimer phase and Haldane phase.
It is known that this critical theory is described 
as $c=\frac{3}{2}$ CFT or 
$k=2$ SU(2) WZW model \cite{kn},
which is equivalent to three Majorana fermions.

For general spin s, Takhtajan-Babujian type solvable models
are described as Wess-Zumino-Wittten non-linear $\sigma$ model(WZW
model)
with topological coupling constant $k=2s$ \cite{ah}, \cite{agsz},
in terms of the analysis for quantum spin systems 
using non-Abelian bosonization \cite{nabwzw}.

Another Bethe Ansatz solvable point 
at $\theta = \frac{\pi}{4}$
is equivalent to 
$k=1$ SU(3) WZW model with $c=2$.
This is known as Uimin-Lai-Sutherland model \cite{uls}.
Above $\theta > \frac{\pi}{4}$, 
although there is not SU(3) symmetry,
the massless region extends
with the $Z_3$ symmetry quasi breaking,
and soft modes appear at $q= \pm \frac{2 \pi}{3}$.
Recently from non-Abelian bosonization and renormalization
group analysis, 
it is shown that this is a variant of the BKT transition
(but the different universality class \cite{ik}).

\section{Wess-Zumino-Witten model (Kac-Moody algebra) \cite{gins},\cite{agsz}}

The minimal conformal field theory with the smallest spectrum
containing currents obeying the Kac-Moody algebra with 
central charge $c$ is the Wess-Zumino-Witten non-linear $\sigma$
model, with topological coupling constant $k$.
Thus the relation between $c$ and $k$ is 
\begin{equation}
 c=\frac{k \dim G}{k+\frac{C_A}{2}},
\end{equation}
where $\dim G$ is dimension for an arbitrary representation 
and $\frac{C_A}{2}$ is called as the dual Coxeter number.
In the present case, since the system has SU(2) symmetry,
we have $\dim G = 3$, and $\frac{C_A}{2} = 2$.

Now we think of the case with the SU(2) symmetry.
In this case, primary fields can be classified according to 
their left and right moving spin. There are operators with 
$s_L=s_R=0, \frac{1}{2},...,\frac{k}{2}$.
Their scaling dimensions are 
\begin{equation}
 x=\frac{2s_L \left( s_L + 1 \right)}{2+k}.
\end{equation}
The operators with $s_L$ half-odd integer (integer) are odd (even)
under translation by one site; thus they correspond to states
with momentum $\pi$ (zero).

Here we think of the case of the $k=2$ SU(2) WZW model.
In the case of $s_L=s_R=\frac{1}{2}$,
which forms $\frac{1}{2} \otimes \frac{1}{2} = 1 \oplus 0$,
the scaling dimension is $x=\frac{3}{8}$.
In the case of $s_L=s_R=1$ which forms
$1 \otimes 1 = 0 \oplus 1 \oplus 2$,
the scaling dimension is $x=1$.

Besides Kac-Moody primary operators,
there is a marginal operator for all $k$,
namely $ \vec{J}_L \cdot \vec{J}_R$
(which is a primary field with respect to the Virasoro algebra).
Current operator $\vec{J}_L$ 
has the conformal weight $(h,\bar{h})=(1,0)$ and
$\vec{J}_R$ has the conformal weight $(h,\bar{h})=(0,1)$.
Thus the $ \vec{J}_L \cdot \vec{J}_R$ operator has the conformal weight 
$(h,\bar{h})=(1,1)$
which has the scaling dimension $x=h+\bar{h}=2$
and the conformal spin 0, i.e. wave number $q=0$.

Current operators $ \vec{J}_L$, $\vec{J}_R$ themselves
have conformal spin $ \pm 1$, corresponding to the wave number
$q= \pm 2 \pi/L$, thus they are related with the spin wave velocity.

\section{logarithmic correction to scaling dimension (or energy gap)\cite{agsz}}

According to the non-Abelian bosonization, 
in SU(2) symmetric gapless system,
the excitation energies (for $q= \pi$ states), 
including logarithmic corrections, are 
\begin{equation}
 \Delta \mbox{E} = \mbox{E}_i - \mbox{E}_0 \approx \frac{2 \pi v}{L}
  \left( x_i - \frac{ \left\langle \vec{S}_L \cdot \vec{S}_R \right\rangle}{ \ln L}
  \right).
\end{equation}
where $L$ is a system size, and $v$ is a spin wave velocity.
Here $\vec{S}=\vec{S}_L+\vec{S}_R$ is total spin which is a
conserved quantity.
Note that
\begin{eqnarray}
 \vec{S}_L \cdot \vec{S}_R&=& \frac{1}{2} \left( \vec{S}_L + \vec{S}_R \right)^2
                             -\frac{1}{2} \vec{S}_L^2 - \frac{1}{2} \vec{S}_R^2 \\
                           &=&\frac{1}{2} s(s+1) - \frac{1}{2} s_L \left(s_L +1 \right) 
                                 - \frac{1}{2} s_R \left(s_R +1 \right).
\end{eqnarray}
In the case of $s_L=s_R=\frac{1}{2}$,
since $s_L \otimes s_R$=$\frac{1}{2} \otimes \frac{1}{2}$=$0 \oplus 1$,
it forms a triplet s=1 and a singlet s=0.
The states $s_L=s_R=\frac{1}{2}$ correspond to a wave vector $q=\pi$.
For s=1(triplet) case,
we obtain
\begin{equation}
 \left\langle triplet \mid \vec{S}_L \cdot \vec{S}_R \mid triplet \right\rangle = \frac{1}{4}.
\end{equation}
On the other hand, for s=0(singlet) case,
\begin{equation}
 \left\langle singlet \mid \vec{S}_L \cdot \vec{S}_R \mid singlet \right\rangle = - \frac{3}{4}.
\end{equation}
Therefore we obtain 
\begin{equation}
 \Delta \mbox{E}(s=1) =
 \mbox{E}_i - \mbox{E}_0 \approx \frac{2 \pi v}{L}
 \left( x_i - \frac{1}{4} \frac{1}{ \ln L}
 \right),
\end{equation}
and
\begin{equation}
 \Delta \mbox{E} (s=0) =
 \mbox{E}_i - \mbox{E}_0 \approx \frac{2 \pi v}{L}
 \left( x_i +   \frac{3}{4} \frac{1}{ \ln L}
 \right),
\end{equation}
thus we can remove logarithmic corrections of energy gap.
\begin{equation}
 x_i = \frac{ L }{8 \pi v} \left[ 3 \Delta \mbox{E}(s=1)
                                + \Delta \mbox{E}(s=0) \right].
\end{equation}

\end{document}